\documentclass[10pt, final, twocolumn, conference, letterpaper, twoside]{IEEEtran}
\usepackage{epsfig}
\usepackage{graphicx}
\usepackage{subfigure}

\newtheorem{theorem}{Theorem}

\newtheorem{definition}{Definition}

\usepackage{amssymb}
\usepackage{makeidx}
\usepackage{amsmath}
\usepackage{graphicx}
\usepackage{epsf}
\usepackage{epsfig}
\usepackage{ccaption}
\usepackage{array}
\usepackage{tabularx}
\usepackage{multirow}
\usepackage{epsfig}
\usepackage{cite}

\begin{document}

\title{\LARGE{Coalition Games with Cooperative Transmission:
A Cure for the Curse of Boundary Nodes in Selfish
Packet-Forwarding Wireless Networks}}
\author{Zhu Han$^*$ and H. Vincent Poor$^+$\\
\vspace{2mm} $^*$Department of Electrical and Computer
Engineering,
\\Boise State
University, Idaho, USA\\
\vspace{2mm} $^+$Department of Electrical Engineering, \\
Princeton University, New Jersey, USA   \thanks{This research was
supported by the National Science Foundation under Grants
ANI-03-38807 and CCR-02-05214, and by the Defense Advanced
Research Projects Agency under Grant HR001-06-1-0052.}}

\maketitle\pagestyle{empty}

\begin{abstract}

In wireless packet-forwarding networks with selfish nodes,
applications of a repeated game can induce the nodes to forward
each others' packets, so that the network performance can be
improved. However, the nodes on the boundary of such networks
cannot benefit from this strategy, as the other nodes do not
depend on them. This problem is sometimes known as {\em the curse
of the boundary nodes}. To overcome this problem, an approach
based on coalition games is proposed, in which the boundary nodes
can use cooperative transmission to help the backbone nodes in the
middle of the network. In return, the backbone nodes are willing
to forward the boundary nodes' packets. The stability of the
coalitions is studied using the concept of a core. Then two types
of fairness, namely, the min-max fairness using nucleolus and the
average fairness using the Shapley function are investigated.
Finally, a protocol is designed using both repeated games and
coalition games. Simulation results show how boundary nodes and
backbone nodes form coalitions together according to different
fairness criteria. The proposed protocol can improve the network
connectivity by about 50\%, compared with pure repeated game
schemes.

\end{abstract}

% ------------------------------ SECTION ----------------------------------
\section{Introduction}\label{sec:intro}

In wireless networks with selfish nodes such as ad hoc networks,
the nodes may not be willing to fully cooperate to accomplish the
overall network goals. Specifically  for the packet-forwarding
problem, forwarding of other nodes' packets consumes a node's
limited battery energy. Therefore, it may not be in a node's best
interest to forward other's arriving packets. However, refusal to
forward other's packets non-cooperatively will severely affect the
network functionality and thereby impair a node's own performance.
Hence, it is crucial to design a mechanism to enforce cooperation
for packet forwarding among greedy and distributed nodes.

The packet-forwarding problem in ad hoc networks has been
extensively studied in the literature. The fact that nodes act
selfishly to optimize their own performance has motivated many
researchers to apply game theory \cite{Game_theory1,Game_theory2}
in solving this problem. Broadly speaking, the approaches used to
encourage packet-forwarding can be categorized into two general
types. The first type  makes use of virtual payments. Pricing
\cite{Crowfort_Gibbens_Kelly_Ostring02} and credit based method
\cite{Zhong_Chen_Yang03} fall into this first type. The second
type of approach is related to personal and community enforcement
to maintain the long-term relationship among nodes. Cooperation is
sustained because defection against one node causes personal
retaliation or sanction by others. \emph{Watchdog} and
\emph{pathrater} are proposed in \cite{Marti_Giuli_Kai_Baker00} to
identify misbehaving nodes and deflect traffic around them.
Reputation-based protocols are proposed in
\cite{Buchegger_LeBoudec02} and \cite{Michiardi_Molva03}. In
\cite{Altman_Kherani_Michiardi_Molva05}, a model is considered to
show cooperation among participating nodes. The packet forwarding
schemes using ``TIT for TAT" schemes are proposed in
\cite{infocom03}. In \cite{hanzhu2}, a cartel maintenance
framework is constructed for distributed rate control for wireless
networks. In \cite{hanzhu1}, self-learning repeated game
approaches are constructed to enforce cooperation and to study
better cooperation. Some recent works for game theory to enhance
energy-efficient behavior in infrastructure networks can be found
in \cite{Meshkati1,Meshkati2,Meshkati3,Meshkati4}.

However, for packet-forwarding networks, there exists the
so-called {\em curse of boundary nodes}. The nodes at the boundary
of the network must depend on the backbone nodes in the middle of
the networks to forward their packets. On the other hand, the
backbone nodes will not depend on the boundary nodes. As a result,
the backbone nodes do not worry about retaliation or lost
reputation for not forwarding the packets of the boundary nodes.
This fact causes the curse of the boundary nodes. In order to cure
this curse, in this paper, we propose an approach based on
cooperative game coalitions using cooperative transmission.

Recently, cooperative transmission \cite{bib:Aazhang1}
\cite{bib:Laneman2} has gained considerable attention as a
transmit strategy for future wireless networks. The basic idea of
cooperative transmission is that the relay nodes can help the
source node's transmission by relaying a replica of the source's
information. Cooperative communications efficiently takes
advantage of the broadcast nature of wireless networks, while
exploiting the inherent spatial and multiuser diversities. The
energy-efficient broadcast problem in wireless networks is
considered in \cite{Yates2}. The work in \cite{Luo} evaluates the
cooperative diversity performance when the best relay is chosen
according to the average SNR, and the outage probability of relay
selection based on instantaneous SNRs. In \cite{Bletsas}, the
authors propose a distributed relay selection scheme that requires
limited network knowledge with instantaneous SNRs. In
\cite{bib:zhuAhmed}, the relay assignment problem is solved for
multiuser cooperative communications. In
\cite{bib:zhuwhohelpswhom}, cooperative resource allocation for
OFDM is studied. A game theoretic approach for relay selection has
been proposed in \cite{globecom_zhu}. In \cite{bib:Adve}, the
centralized power allocation schemes are presented by assuming all
the relay nodes helped. In \cite{bib:Madsen}, cooperative routing
protocols are constructed based on non-cooperative routes.

Using cooperative transmission, boundary nodes can serve as relays
and provide some transmission benefits for the backbone nodes that
can be viewed as source nodes. In return, the boundary nodes are
rewarded for packet-forwarding. To analyze the benefits and
rewards, we investigate a game coalition that describes how much
collective payoff a set of nodes can gain and how to divide the
payoff. We investigate the stability and payoff division using
concepts such as the core, nucleolus, and Shapley function. Two
types of fairness are defined, namely, the min-max fairness using
nucleolus and average fairness using the Shapley function. Then,
we construct a protocol using both repeated games and coalition
games. From the simulation results, we investigate how boundary
nodes and backbone nodes form coalitions according to different
fairness criteria. The proposed protocol can improve the network
connectivity by about 50\%, compared to the pure repeated game
approach.

This paper is organized as follows: In Section \ref{sec:model},
repeated game approaches are reviewed and the curse of the
boundary nodes is explained. In Section \ref{sec:protocol}, the
cooperative transmission model is illustrated and the
corresponding coalition games are constructed. Stability and two
types of fairness are investigated. A protocol that exploits the
properties of our approach is also proposed. Simulation results
are shown in Section \ref{sec:simulation} and conclusions are
given in Section \ref{sec:conclusion}.

% ------------------------------ SECTION ----------------------------------
\section{Repeated Games and Curse of Boundary Nodes\label{sec:model}}

A wireless packet-forwarding network can be modeled as a directed
graph $G(L,A)$, where $L$ is the set of all nodes and $A$ is the
set of all directed links $(i,l), i,l\in L$. Each node $i$ has
several transmission destinations which are included in set $D_i$.
To reach the destination $j$ in $D_i$, the available routes form a
{\em depending graph} $G_i^j$ whose nodes represents the potential
packet-forwarding nodes. The transmission from node $i$ to node
$j$ depends on a subsect of the nodes in $G_i^j$ for
packet-forwarding. Notice that this dependency can be mutual. One
node depends on the other node, while the other node can depend on
this node as well. In general, this mutual dependency is common,
especially for backbone nodes at the center of the network. In the
remainder of this section, we will discuss how to make use of this
mutual dependency for packet-forwarding using a repeated game, and
then we will explain the curse of boundary nodes.

\subsection{Repeated Games for Mutually Dependent Nodes}

A repeated game is a special type of dynamic game (a game that is
played multiple times). When the nodes interact by playing a
similar static game (which is played only once) numerous times,
the game is called a repeated game. Unlike a static game, a
repeated game allows a strategy to be contingent on the past
moves, thus allowing reputation effects and retribution, which
give possibilities for cooperation. The game is defined as
follows:

\begin{definition}
A $T$-period {\em repeated game} is a dynamic game in which, at
each period $t$, the moves during periods $1,\dots, t-1$ are known
to every node. In such a game, the total discounted payoff for
each node is computed by $ \sum_{t=1}^T \beta^{t-1} u_i(t), $
where $u_i(t)$ denotes the payoff to node $i$ at period $t$ and
where $\beta$ is a discount factor. Note that $\beta$ represents
the node's patience or on the other hand how important the past
affects the current payoff. If $T = \infty$, the game is referred
as an infinitely-repeated game. The average payoff to node $i$ is
then given by:
\begin{equation}
    u_i=(1-\beta)\sum_{t=1}^{\infty}\beta ^{t-1}u_i(t).
    \label{payoff}
\end{equation}
\end{definition}

It is known that repeated games can be used to induce greedy nodes
in communication networks to show cooperation. In
packet-forwarding networks, if a greedy node does not forward the
packets of other nodes, it can enjoy benefits such as power
saving. However, this node will get punishment from the other
nodes in the future if it depends on the other nodes to forward
its own packets. The benefit of greediness in the short term will
be offset by the loss associated with punishment in the future. So
the nodes will rather act cooperatively if the nodes are
sufficiently patient. From the Folk theorem below, we infer that
in an infinitely repeated game, any feasible outcome that gives
each node a better payoff than the Nash equilibrium
\cite{Game_theory1,Game_theory2} can be obtained.

\begin{theorem}
({\em Folk Theorem} \cite{Game_theory1,Game_theory2}) Let
($\hat{u}_1, \dots , \hat{u}_L$) be the set of payoffs from a Nash
equilibrium and let ($u_1, \dots , u_L$) be any feasible set of
payoffs. There exists an equilibrium of the infinitely repeated
game that attains any feasible solution ($u_1, \dots , u_L$) with
$u_i \geq \hat{u}_i, \forall i$ as the average payoff, provided
that $\beta$ is sufficiently close to 1.
\end{theorem}

In the literature of packet-forwarding wireless networks, the
conclusion of the above Folk theorem is achieved by several
approaches. Tit-for-tat \cite{Altman_Kherani_Michiardi_Molva05}
\cite{infocom03} is proposed so that all mutually dependent nodes
have the same set of actions. A cartel maintenance scheme
\cite{hanzhu2} has closed-form optimal solutions for both
cooperation and non-cooperation. A self-learning repeated game
approach is proposed in \cite{hanzhu1} for individual distributed
nodes to study the cooperation points and to develop protocols for
maintaining them. Given the previous attention to the problem of
nodes having mutual dependency, we will assume in this paper that
the packet-forwarding problem of selfish nodes with mutual
dependency has been solved and we will focus instead on the
problems encountered by the boundary nodes.

%\begin{figure}[ht]
%\centerline{
%\begin{tabular}{cc}
%\psfig{system_model.eps, width=75truemm}&\psfig{figure=Game_model.eps,width=75truemm}\\
%{\small (a)}& {\small (b)}
%\end{tabular}
%}
%\caption{\label{example}{(a) Example of the Curse of Boundary
%Nodes (b) Coalition Game Model with Cooperative
%Transmission}}\vspace{-5mm}
%\end{figure}

%\begin{figure}\label{example}
%  \centering
%      \subfigure[Example of the Curse of Boundary
%Nodes] {
%        \includegraphics[width=65truemm]{system_model.eps}}
%        \hspace{10truemm}%
%        \subfigure[Coalition Game Model with Cooperative Transmission] {
%        \includegraphics[width=65truemm]{Game_model.eps}}
%    %  \caption{Multiple Relay Case and Performance Comparison}
%    % \label{fig: multiple}
%    \vspace{-10mm}
%\end{figure}

%\begin{figure}[htbp]
%\begin{center}
%    \epsfig{file=system_model.eps,width=80truemm}
%\end{center}
%\caption{Example of the Curse of Boundary
%Nodes}\label{curse_example}\vspace{-5mm}
%\end{figure}

\begin{figure}
    \centering
    \includegraphics[width=80truemm]{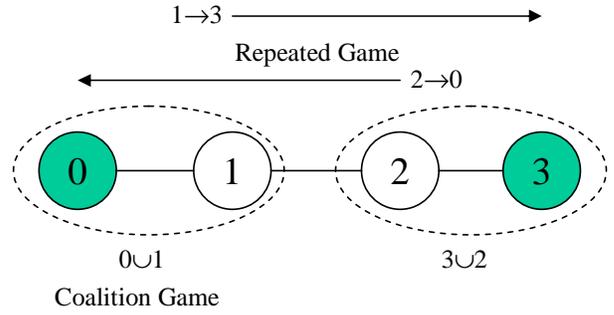}
    \caption{\footnotesize{Example of the Curse of Boundary Nodes}}
    \label{example}
\end{figure}

\subsection{Curse of Boundary Nodes\label{curse}}

When there is no mutual dependency, the  curse of boundary nodes
occurs, an example of which is shown in Figure \ref{example}.
Suppose node $1$ needs to send data to node $3,$ and node $2$
needs to send data to node $0$. Because node $1$ and node $2$
depend on each other for packet-forwarding, they are obliged to do
so because of the possible threat or retaliation from the other
node. However, if node $0$ wants to transmit to node $2$ and node
$3$, or node $3$ tries to communicate with node $0$ and node $1$,
the nodes in the middle have no incentive to forward the packets
due to their greediness. Moreover, this greediness cannot be
punished in the future since the dependency is not mutual. This
problem is especially severe for the nodes on the boundary of the
networks, so it is called {\em the curse of boundary nodes}.

On the other hand, if node $0$ can form a coalition with node $1$
and help node $1$'s transmission (for example to reduce the
transmitted power of node $1$), then node $1$ has the incentive to
help node $0$ transmit as a reward. A similar situation arises
 for node $3$ to form a coalition with node $2$. We call
 nodes like $1$ and $2$ {\it backbone nodes}, while nodes
like $0$ and $3$ are  {\it boundary nodes}. In the following
section, we will study how coalitions can be formed to address
this issue using cooperative transmission.

% ------------------------------ SECTION ----------------------------------
\section{Coalition Games with Cooperative Transmission}\label{sec:protocol}

In this section, we first study a cooperative transmission
technique that allows nodes to participate in coalitions. Then, we
formulate a coalition game with cooperative transmission.
Furthermore, we investigate the fairness issue and propose two
types of fairness definitions. Finally, a protocol for
packet-forwarding using repeated games and coalition games is
constructed.

\subsection{Cooperative Transmission System Model}

First, we discuss the traditional direct transmission case. The
source transmits its information to the destination with power
$P_d$. The received signal to noise ratio (SNR) is
\begin{equation}\label{SNR_direct}
\Gamma_{d}=\frac {P_dh_{s,d}}{\sigma ^2},
\end{equation}
where $h_{s,d}$ is the channel gain from the source to the
destination and $\sigma^2$ is the noise level. To achieve the
minimal link quality $\gamma$, we need for the transmitted power
to be sufficiently large so that $\Gamma_d \geq \gamma$. The
transmitted power is also upper bounded by $P_{max}$.

%\begin{figure}[htbp]
%\begin{center}
%    \epsfig{file=Game_model.eps,width=70truemm}
%\end{center}
%\caption{Coalition Game Model with Cooperative
%Transmission}\label{game_model}\vspace{-5mm}
%\end{figure}

\begin{figure}
    \centering
    \includegraphics[width=80truemm]{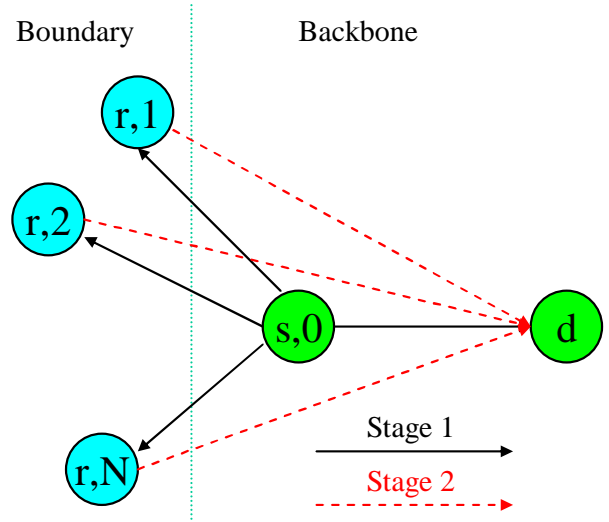}
    \caption{\footnotesize{Coalition Game with Cooperative Transmission}}
    \label{example_ct}
\end{figure}

Next, we consider  multiple nodes using the amplify-and-forward
protocol \cite{bib:Aazhang1}\footnote{Other cooperative
transmission protocols can be exploited in a similar way.} to
transmit in two stages as shown in Figure \ref{example_ct}. In
stage one, the source node (denoted as node $0$) transmits its
information to the destination, and due to the broadcast nature of
the wireless channels, the other nodes can receive the
information. In stage two, the remaining $N$ relay nodes help the
source by amplifying the source signal. In both stages, the source
and the relays transmit their signals through orthogonal channels
using schemes like TDMA, FDMA, or orthogonal CDMA.

In stage one, the source transmits its information and the
received signals at the destination and the relays can be written
respectively as
\begin{equation}
y_{s,d}=\sqrt{P_0}h_{s,d}x+n_{s,d},
\end{equation}
\begin{equation}
\mbox{and  }y_{s,r_i}=\sqrt{P_0}h_{s,r_i}x+n_{s,r_i}, \forall i\in
\{1,\dots, N\},
\end{equation}
where $P_0$ is the transmitted power of the source, $x$ is the
transmitted symbol with unit power, $h_{s,r_i}$ is the channel
gain from the source to relay $i$, and $n_{s,d}$ and $n_{s,r_i}$
are the thermal noise processes at the destination and relay,
respectively. Without significant loss of generality, we assume
that all thermal noises have the same power $\sigma ^2$.

In stage two, each relay amplifies the received signal from the
source and retransmits it to the destination. The received signal at
the destination for relay $i$ can be written as
\begin{equation}
y_{r_i,d}=\frac{\sqrt{P_i}}{\sqrt{P_0|h_{s,r_i}|^2+\sigma^2}}
h_{r_i,d}y_{s,r_i}+n_{r_i,d},
\end{equation}
where $P_i$ is relay $i$'s transmit power, $h_{r_i,d}$ is the
channel gain from relay $i$ to the destination,  and $n_{r_i,d}$
is the thermal noise with variance $\sigma ^2$.

At the destination, the signal received at stage one and the $N$
signals received at stage two are combined using maximal ratio
combining (MRC). The SNR at the output of MRC is
\begin{equation}\label{SNR_coop}
\Gamma=\Gamma_0+\sum _{i=1}^N \Gamma _i,
\end{equation}
where $\Gamma_0=\frac{P_0|h_{s,d}|^2}{\sigma ^2}$ and
\begin{equation}
\Gamma _i =\frac {P_0P_i|h_{s,r_i}|^2|h_{r_i,d}|^2} {\sigma^2
(P_0|h_{s,r_i}|^2+P_i|h_{r_i,d}|^2+\sigma^2)}.
\end{equation}

On comparing (\ref{SNR_coop}) with (\ref{SNR_direct}), in order to
achieve the desired link quality $\gamma$, we can see that the
required power is always less than the direct transmission power,
i.e., $P_0< P_d$. So cooperation transmission can reduce the
transmit power of the source node. This fact can give incentives
of mutual benefits for the backbone nodes (acting as sources) and
the boundary nodes (acting as relays), and consequently can cure
the curse of the boundary nodes mentioned in Section \ref{curse}.

\subsection{Coalition Game Formation for Boundary Nodes}

In this subsection, we study possible coalitions between the
boundary nodes and the backbone nodes, for situations in which the
boundary nodes can help relay the information of the backbone
nodes using cooperative transmission. In the following, we first
define some basic concepts that will be needed in our analysis.

\begin{definition}
A {\em coalition} $S$ is defined to be a subset of the total set of
nodes $\mathbb{N}=\{0,\dots , N\}$. The nodes in a coalition want
to cooperate with each other. The {\em coalition form} of a game
is given by the pair $(\mathbb{N},v)$, where $v$ is a real-valued
function, called  the {\em characteristic function}. $v(S)$ is the
value of the cooperation for coalition $S$ with the following
properties:
\begin{enumerate}
\item $v(\emptyset)=0$.

\item Super-additivity: if $S$ and $Z$ are disjoint coalitions
($S\bigcap Z=\emptyset$), then $v(S)+v(Z)\leq v(S\bigcup Z)$.
\end{enumerate}
\end{definition}

The coalition states the benefit obtained from cooperation
agreements. But we still need to examine whether or not the nodes
are willing to participate in the coalition. A coalition is called
{\it stable} if no other coalition will have the incentive and
power to upset the cooperative agreement. Such division of $v$ is
called a point in the {\em core}, which is defined by the
following definitions.

\begin{definition}
A payoff vector $\textbf U=(U_0,\dots, U_{N})$ is said to be {\em
group rational} or {\em efficient} if $\sum_{i=0}^{N}U_i=
v(\mathbb{N})$. A payoff vector $\textbf U$ is said to be {\em
individually rational} if the node can obtain the benefit no less
than acting alone, i.e. $U_i\geq v(\{i\}),\ \forall i$. An {\em
imputation} is a payoff vector satisfying the above two
conditions.
\end{definition}

\begin{definition}
An imputation $\textbf U$ is said to be unstable through a
coalition $S$ if $v(S)>\sum_{i\in S}U_i$, i.e., the nodes have
incentive for coalition $S$ and upset the proposed $\textbf U$.
The set $C$ of a stable imputation is called the {\em core}, i.e.,
\begin{equation}
C=\{\textbf U:\sum_{i\in \mathbb{N}}U_i=v(\mathbb{N}) \mbox{ and
}\sum_{i\in S}U_i\geq v(S),\ \forall S\in \mathbb{N}\}.
\end{equation}
\end{definition}

%There are two types of coalitions with transferable utility or
%with untransferable utility. The transferable utility is defined
%as:
%\begin{definition}
%Transferable utility is the payoff that can be transferred between
%nodes, like money. For coalition games with transferable utility,
%one node can make a transfer to another so as to get them to
%participate in a coalition. These side payments facilitate
%coalition formation.
%\end{definition}
%For our case, the transferable utility between backbone nodes and
%boundary nodes is equivalent to energy.

In the economics literature, the core gives a reasonable set of
possible shares. A combination of shares is in the core if there
is no sub-coalition in which its members may gain a higher total
outcome than the combination of shares of concern. If a share is
not in the core, some members may be frustrated and may think of
leaving the whole group with some other members and form a smaller
group.

In the packet-forwarding network as shown in Figure
\ref{example_ct}, we first assume one backbone node to be the
source node (node $0$) and the nearby boundary nodes (node $1$ to
node $N$) to be the relay nodes. We will discuss the case of
multiple source nodes later. If no cooperative transmission is
employed, the utilities for the source node and the relay nodes
are
\begin{equation}
v(\{ 0\})=-P_d,\mbox{ and } v(\{ i\})=-\infty, \forall i=1,\dots,
N.
\end{equation}
With cooperative transmission and a grand coalition that includes
all nodes, the utilities for the source node and the relay nodes
are
\begin{equation}\label{alpha}
U_0=-P_0-\sum_{i=1}^N \alpha_i P_d
\end{equation}
\begin{equation}\label{Ui}
\mbox{and }U_i=-\frac{P_i}{\alpha_i},
\end{equation}
where $\alpha_i$ is the ratio of the number of packets that the
backbone node is willing to forward for boundary node $i$, to the
number of packets that the boundary node $i$ relays for the
backbone node using cooperative transmission. Here we use negative
power as the utility so as to be consistent with the conventions
used in the game theory literature. Smaller $\alpha_i$ means the
boundary nodes have to relay more packets before realizing the
rewards of packet forwarding. The other interpretation of the
utility is as the average power per transmission for the boundary
nodes\footnote{Notice that we omit the transmitted power needed to
send the boundary node's own packet to the backbone node, since it
is irrelevant to the coalition.}. The following theorem gives
conditions under which the core is not empty, i.e, in which the
grand coalition is stable.

\begin{theorem}
The core is not empty if $\alpha_i\geq 0,\ i=1,\dots, N$, and
$\alpha_i$ are such that $U_0\geq v(\{ 0\})$, i.e,
\begin{equation}\label{condition_core}
\sum_{i=1}^N \alpha_i \leq \frac{P_d-P_0}{P_d}.
\end{equation}
\end{theorem}
\begin{proof}
First, any relay node will get $-\infty$ utility if it leaves the
coalition with the source node, so no node has incentive to leave
coalition with node $0$. Then, from (\ref{SNR_coop}), the
inclusion of relay nodes will increase the received SNR
monotonically. So $P_0$ will decrease monotonically with the
addition of any relay node. As a result, the source node has the
incentive to include all the relay nodes, as long as the source
power can be reduced, i.e., $U_0\geq v(\{ 0\})$. A grand coalition
is formed and the core is not empty if (\ref{condition_core})
holds.
\end{proof}

The concept of the core defines the stability of a utility
allocation. However, it does not define how to allocate the
utility. For the proposed game, each relay node can obtain
different utilities by using different values of  $\alpha_i$. In
the next two subsections, we study how to achieve min-max fairness
and average fairness.

\subsection{Min-Max Fairness of a Game Coalition using Nucleolus}

We introduce the concepts of {\em excess}, {\em kernel}, and {\em
nucleolus}\cite{Game_theory1,Game_theory2}. For a fixed
characteristic function $v$, an imputation $\textbf U$ is found
such that, for each coalition $S$ and its associated
dissatisfaction, an optimal imputation is calculated to minimize
the maximum dissatisfaction. The dissatisfaction is quantified as
follows.
\begin{definition}
The measure of dissatisfaction of an imputation $\textbf U$
for a coalition $S$ is defined as the {\em excess}:
\begin{equation}
e(\textbf U,S)=v(S)-\sum_{j\in S}U_j.
\end{equation}
\end{definition}
Obviously, any imputation $\textbf U$ is in the core, if and only
if all its excesses are negative or zero.

\begin{definition}
A {\em kernel} of $v$ is the set of all allocations $\textbf U$
such that
\begin{equation}
\max_{S \subseteq \mathbb{N}-j,i\in S}e(\textbf U,S)=\max_{T
\subseteq \mathbb{N}-i,j\in T}e(\textbf U,T).
\end{equation}
If nodes $i$ and $j$ are in the same coalition, then the highest
excess that $i$ can make in a coalition without $j$ is equal to
the highest excess that $j$ can make in a coalition without $i$.
\end{definition}

\begin{definition}
The {\em nucleolus} of a game is the allocation $\textbf U$ that
minimizes the maximum excess:
\begin{equation}
    \textbf U=\arg \min _{\textbf U} (\max\ e(\textbf U,S),\ \forall S).
\end{equation}
\end{definition}

The nucleolus of a game has the following property: The nucleolus
of a game in coalitional form exists and is unique. The nucleolus
is group rational and individually rational. If the core is not
empty, the nucleolus is in the core and kernel. In other word, the
nucleolus is the best allocation under the min-max criterion.

Using the above concepts, we prove the following theorem to show
the optimal $\alpha_i$ in (\ref{alpha}) to have min-max fairness.
\begin{theorem}
The maximal $\alpha_i$ to yield the nucleolus of the proposed
coalition game is given by
\begin{equation}\label{minmax_solution}
\alpha_i=\frac{P_d-P_0(\mathbb{N})}{NP_d},
\end{equation}
where $P_0(\mathbb{N})$ is the required transmitted power of the
source when all relays transmit with transmitted power $P_{max}$.
\end{theorem}
\begin{proof}
Since for any coalition other than the grand coalition, the excess
will be $-\infty$, we need only consider the grand coalition.
Suppose the min-max utility is $\mu$ for all nodes, i.e.
\begin{equation}
\mu=-\frac {P_i}{\alpha_i}.
\end{equation}
From (\ref{condition_core}) and since $U_i$ is monotonically
increasing with $\alpha_i$ in (\ref{Ui}), we have
\begin{equation}
{\alpha_i} = \frac{P_i}{\sum_{i=1}^N{P_i}}\cdot
\frac{(P_d-P_0)}{P_d}.
\end{equation}
Since $P_0$ in (\ref{SNR_coop}) is a monotonically increasing
function of $P_i$, to achieve the maximal $\alpha_i$ and $\mu$,
each relay transmits with the largest possible power $P_{max}$.
Notice here we assume the backbone node can accept arbitrarily
small power gain to join the coalition.
\end{proof}

\subsection{Average Fairness of Game Coalition using the Shapley Function}

The core concept defines the stability of an allocation of payoff
and the nucleolus concept quantifies the min-max fairness of a
game coalition. In this subsection, we study another average
measure of fairness for each individual using the concept of a
Shapley function \cite{Game_theory1,Game_theory2}.

\begin{definition}
A {\em Shapley function} $\phi$ is a function that assigns to each
possible characteristic function $v$ a vector of real numbers,
i.e.,
\begin{equation}
    \phi (v)=(\phi_0(v),\phi_1(v),\phi_2(v),\dots, \phi_N(v))
\end{equation}
where $\phi_i(v)$ represents the worth or value of node $i$ in the
game. There are four Shapley Axioms that $\phi(v)$ must satisfy
\begin{enumerate}
\item {\em Efficiency Axiom}: $\sum_{i\in
\mathbb{N}}\Phi_i(v)=v(\mathbb{N})$.

\item {\em Symmetry Axiom}: If node $i$ and node $j$ are such that
$v(S\bigcup \{i\})=v(S\bigcup \{j\})$ for every coalition $S$ not
containing node $i$ and node $j$, then $\phi_i(v)=\phi_j(v)$.

\item {\em Dummy Axiom}: If node $i$ is such that $v(S)=v(S\bigcup
\{ i\})$ for every coalition $S$ not containing $i$, then
$\phi_i(v)=0$.

\item {\em Additivity Axiom}: If $u$ and $v$ are characteristic
functions, then $\phi(u+v)=\phi(v+u)=\phi(u)+\phi(v)$.

\end{enumerate}

It can be proved that there exists a unique function $\phi$
satisfying the Shapley axioms. Moreover, the Shapley function can
be calculated as
\begin{equation}
\phi_i(v)=\sum_{S\subset
\mathbb{N}-i}\frac{(|S|)!(N-|S|)!}{(N+1)!}[v(S\cup \{ i\})-v(S)].
\end{equation}
Here $|S|$ denotes the size of set $S$ and $\mathbb{N}=\{0, 1,
\dots N\}$.

\end{definition}

The physical meaning of the Shapley function can be interpreted as
follows. Suppose one backbone node plus $N$ boundary nodes form a
coalition. Each node joins the coalition in random order. So there
are $(N+1)!$ different ways that the nodes might be ordered in
joining the coalition. For any set $S$ that does not contain node
$i$, there are $|S|!(N-|S|)!$ different ways to order the nodes so
that $S$ is the set of nodes who enter the coalition before node
$i$. Thus, if the various orderings are equally likely,
$|S|!(N-|S|)!/(N+1)!$ is the probability that, when node $i$
enters the coalition, the coalition of $S$ is already formed. When
node $i$ finds $S$ ahead of it as it joins the coalition, then its
marginal contribution to the worth of the coalition is $v(S\cup \{
i\})-v(S)$. Thus, under the assumption of randomly-ordered
joining, the Shapley function of each node is its expected
marginal contribution when it joins the coalition.

In our specific case, we consider the case in which the backbone
node is always in the coalition, and the boundary nodes randomly
join the coalition. We have $v(\{0\})=-P_d$ and
\begin{equation}
    v(\mathbb{N})=P_d-P_0(\mathbb{N})-\sum_{i\in \mathbb{N}} \alpha_i P_d,
\end{equation}
which is the overall power saving. The problem here is how to find
a given node's $\alpha_i$ that satisfies the average fairness,
which is addressed by the following theorem.

\begin{theorem}
The maximal $\alpha_i$ that satisfies the average fairness with
the physical meaning of the Shapley function is given by
\begin{equation}\label{shapley_solution}
\alpha_i=\frac{P_i^s}{P_d},
\end{equation}
where $P_i^s$ is the average power saving with random entering
orders, which is defined as
\begin{eqnarray}
P_i^s=\frac{1}{N}[P_d-P_0(\{i\})]\nonumber \\
+\frac{\sum_{j=1,j\neq i}^N
[P_0(\{j\})- P_0(\{i,j\})]}{N(N-1)} +\cdots.
\end{eqnarray}
\end{theorem}

\begin{proof}
The maximal $\alpha_i$ is solved by the following equations:
\begin{equation}\label{Shapley_eqn}
\left\{
\begin{array}{l}
\frac{\alpha_i}{\alpha_j}=\frac{\phi_i}{\phi_j},\\
v(\mathbb{N})\geq 0.\\
\end{array}
\right.
\end{equation}
The first equation in (\ref{Shapley_eqn}) is the average fairness
according to the Shapley function, and the second equation in
(\ref{Shapley_eqn}) is the condition for a non-empty core. Similar
to min-max fairness, we assume that the backbone node can accept
arbitrarily small power gain to join the coalition.

If boundary node $i$ is the first to join the coalition, the
marginal contribution for power saving is $
\frac{1}{N}[P_d-P_0(\{i\})-\alpha_i P_d]$, where $\frac 1 N$ is
the probability. If boundary node $i$ is the second to join the
coalition, the marginal contribution is $\frac{\sum_{j=1,j\neq
i}^N [P_0(\{j\})+\alpha_j P_d- P_0(\{i,j\})- (\alpha_i+\alpha_j)
P_d]}{N(N-1)}$. By means of some simple derivations, we can obtain
the Shapley function $\phi_i$ as
\begin{eqnarray}\label{phii}
\phi_i= -\alpha_i P_d+ \frac{1}{N}[P_d-P_0(\{i\})]\\
+\frac{\sum_{j=1,j\neq i}^N [P_0(\{j\})- P_0(\{i,j\})]}{N(N-1)}
+\cdots,\nonumber
\end{eqnarray}
and then we can obtain
\begin{equation}
\alpha_i=\frac{[P_d-P_0(\mathbb{N})]P_i^s}{P_d\sum_{j=1}^N P_j^s}.
\end{equation}
Since
\begin{equation}
P_d-P_0(\mathbb{N})=\sum_{j=1}^N P_j^s,
\end{equation}
we prove (\ref{shapley_solution}).
\end{proof}

Notice that different nodes have different values of $P_i^s$, due
to their channel conditions and abilities to reduce the backbone
node's power. Compared with the min-max fairness in the previous
subsection, the average fairness using the Shapley function gives
different nodes different values of $\alpha_i$ according to their
locations.

%\begin{eqnarray}\label{phi0}
%\phi_0\approx -\frac{P_d}{N+1}-\frac{\sum_{i=1}^N i C_0}{N+1}=
%-\frac{P_d}{N+1}-\frac{NC_0}{2}\approx -\frac{NC_0}{2}
%\end{eqnarray}
%where the first term in the right hand side (RHS) is the utility
%without any boundary node, the second term in RHS represents the
%probability times the marginal contribution when the backbone node
%is the second node to join the coalition, the third term in RHS is
%for the case where two boundary nodes are already in the
%coalition, and so on until the size of $S$ is equal to $N$.
%
%If boundary node $i$ enters the coalition with $K$ nodes and finds
%the backbone node is not in the coalition, the marginal
%contribution is zero. If the boundary node is the first to join
%the coalition, the contribution is also zero. The Shapley values
%for the boundary nodes are given by
%\begin{eqnarray}\label{phii}
%\phi_i= -\alpha_i P_d+ \frac{1}{N}[P_d-P_0(\{i\})]
%+\frac{\sum_{j=1,j\neq i}^N [P_0(\{j\})- P_0(\{i,j\})]}{N(N-1)}
%+\cdots.
%\end{eqnarray}
%Here the first term in RHS is the marginal contribution when the
%first node to join the coalition is the backbone node and the
%second is node $i$. The second term in RHS is when $K=2$ and
%$\frac 2 {N+1}$ is the probability that the backbone node join the
%coalition in the first two rounds. In the above Shapley value
%derivation, we have the assumption that the core not empty.

%From (\ref{phi0}) and (\ref{phii}),
%
%\begin{equation}
%\frac {\phi_0}{\phi_i}=\frac{-P_0-\sum_{i=1}^N \alpha_i P_d}{\frac
%{P_{max}}{\alpha_i}}, \forall i.
%\end{equation}

\subsection{Joint Repeated-Game and Coalition-Game Packet-Forwarding Protocol}

Using the above analysis, we now develop a packet-forwarding
protocol based on repeated games and coalition games based on the
following steps.

\begin{center}{{\em Packet-Forwarding Protocol using Repeated
Games and Coalition Games}}
\end{center}
\begin{enumerate}
\item  Route discovery for all nodes.

\item  Packet-forwarding enforcement for the backbone nodes, using
threat of future punishment in the repeated games.

\item Neighbor discovery for the boundary nodes.

\item Coalition game formation.

\item Packet relay for the backbone nodes with cooperative
  transmission.

\item Transmission of the boundary nodes' own packets to the
backbone nodes for forwarding.
 \vspace{5mm}
\end{enumerate}

First, all nodes in the network undergo route discovery. Then each
node knows who depends on it and on whom it depends for
transmission. Using this route information, the repeated games can
be formulated for the backbone nodes. The backbone nodes forward
the other nodes' information because of the threat of future
punishment if these packets are not forwarded. Due to the network
topology, some nodes' transmissions depend on the others while the
others do not depend on these nodes. These nodes are most often
located at the boundary of the network. In the next step, these
boundary nodes try to find their neighboring backbone nodes. Then,
the boundary nodes try to form coalitions with the backbone nodes,
so that the boundary nodes can be rewarded for transmitting their
own packets. Cooperative transmission gives an opportunity for the
boundary nodes to pay some ``credits" first to the backbone nodes
for the rewards of packet-forwarding in return. On the other hand,
competition among the backbone nodes prevents the boundary nodes
from being forced to accept the minimal payoffs.

%\begin{table}
%\caption{Packet-Forwarding Protocol}
%\begin{center}\label{protocol}
%\begin{tabular}{|l|}
%  \hline
%  % after \\: \hline or \cline{col1-col2} \cline{col3-col4} ...
%  1. Route discovery for all nodes.\\
%  \hline
%  2. Packet-forwarding enforcement for the backbone nodes, using threat of future punishment in the repeated games. \\
%  \hline
%  3. Neighbor discovery for the boundary nodes.\\
%  \hline
%  4. Coalition game formation.\\
%  \hline
%  5. Relaying the backbone node's packets with cooperative
%  transmission.\\
%  \hline
%  6. Transmitting the boundary nodes' own information to the backbone nodes for
%  forwarding.\\
%  \hline
%\end{tabular}
%\end{center}
%\end{table}

%\begin{figure}[htbp]
%\begin{center}
%    \epsfig{file=a_R.eps,width=70truemm}
%\end{center}
%\caption{$\alpha$ for Different Channels and No. of
%Nodes}\label{a_R}\vspace{-5mm}
%\end{figure}

%\begin{figure}[htbp]
%\begin{center}
%    \epsfig{file=prob_size.eps,width=70truemm}
%\end{center}
%\caption{Network Connectivity as a function of Network
%Size}\label{prob_size}\vspace{-5mm}
%\end{figure}
% ------------------------------ SECTION ----------------------------------
\section{Simulation Results}\label{sec:simulation}

We model all channels as additive white Gaussian noise channels
having a propagation factor of $3$; that is, power falls off
spatially according to an inverse-cubic law. The maximal
transmitted power is $10$dbmW and the thermal noise level is
-$60$dbmW. The minimal SNR $\gamma$ is 10dB. In the first setup,
we assume the backbone node is located at $(0m,0m)$, and the
destination is located at either $(100m,0m)$ or $(50m,0m)$. The
boundary nodes are located on an arc with angles randomly
distributed from $0.5\pi$ to $1.5\pi$ and with distances varying
from $5$m to $100$m.

\begin{figure}
    \centering
    \includegraphics[width=80truemm]{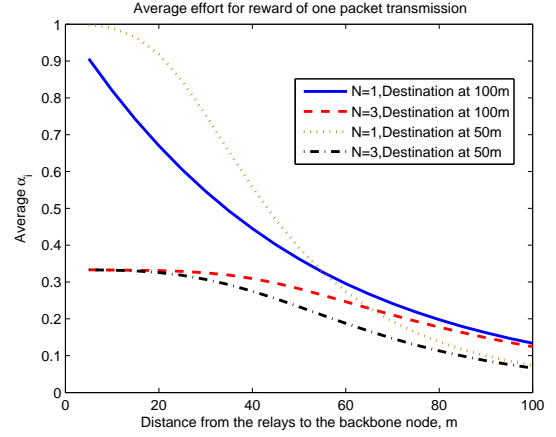}
    \caption{\footnotesize{$\alpha$ for Different Channels and No. of Nodes, Min-Max Fairness}}
    \label{a_R}
\end{figure}

\begin{figure}
    \centering
    \includegraphics[width=80truemm]{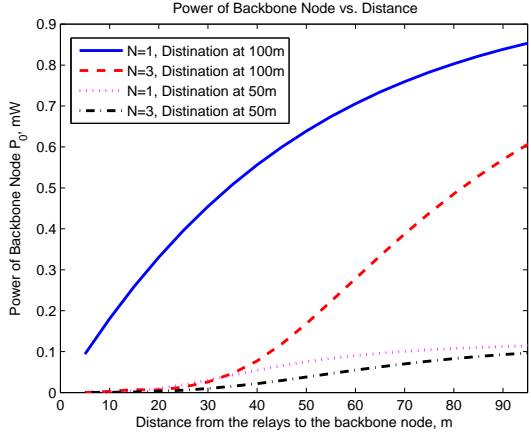}
    \caption{\footnotesize{$P_0$ for Different Channels and No. of Nodes, Min-Max Fairness}}
    \label{P0_R}
\end{figure}

In Figure \ref{a_R}, we study the min-max fairness and show the
average $\alpha_i$ over $1000$ iterations as a function of
distance from the relays to the source node. Due to the min-max
nature, all boundary nodes have the same $\alpha_i$. When the
distance is small, i.e., when the relays are located close to the
source, $\alpha_i$ approaches $\frac 1 N$. This is because the
relays can serve as a virtual antenna for the source, and the
source needs very low power for transmission to the relays. When
the distance is large, the relays are less effective and
$\alpha_i$ decreases, which means that the relays must transmit
more packets for the source to earn the rewards of
packet-forwarding. When the destination is located at $50$m, the
source-destination channel is better than that at 100$m$. When
$N=1$ and the source-destination distance is $50$m, the relays
close to the source have larger $\alpha_i$ and the relays farther
away have lower  $\alpha_i$ than that in the $100$m case. In
Figure \ref{P0_R}, we show the corresponding $P_0$ for the
backbone node. We can see that $P_0$ increases when the distances
between the boundary nodes to the backbone node increase.

If we consider the multiple backbone (multiple core) case  with
min-max fairness, Figure \ref{a_R} and Figure \ref{P0_R} provide
the boundary nodes a guideline for selecting a backbone node with
which to form a coalition. First, a less crowded coalition is
preferred. Second, the nearest backbone node is preferred. Third,
for  $N=1$, if the source-destination channel is good, the closer
backbone node is preferred; otherwise, the farther one can provide
larger $\alpha_i$.

%\begin{figure}
%  \centering
%      \subfigure[$\alpha$ for Different Channels and No. of
%Nodes] {
%        \includegraphics[width=65truemm]{a_R.eps}}
%        \hspace{10truemm}%
%        \subfigure[Network Connectivity as a function of Network
%Size] {
%        \includegraphics[width=65truemm]{prob_size.eps}}
%    %  \caption{Multiple Relay Case and Performance Comparison}
%    % \label{fig: multiple}
%    \vspace{-10mm}\label{a_R}
%\end{figure}

\begin{figure}
    \centering
    \includegraphics[width=80truemm]{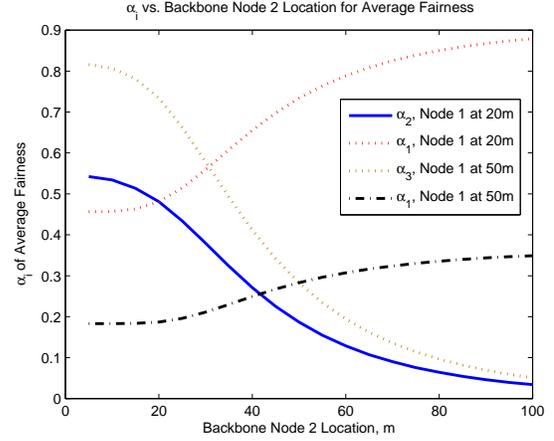}
    \caption{\footnotesize{$\alpha_i$ of Average Fairness for Different Users' Locations}}
    \label{a_shapley}
\end{figure}

Next, we investigate the average fairness using the Shapley
function. The simulation setup is as follows. The backbone node is
located at $(0m,0m)$ and the destination is located at
$(-50m,0m)$. Boundary node $1$ is located at $(20m,0m)$ and
$(50m,0m)$, respectively. Boundary node $2$ moves from $(5m,0m)$
to $(100m,0m)$. The remaining simulation parameters are the same.
In Figure \ref{a_shapley}, we show maximal $\alpha_i$ for two
boundary nodes. We can see that when boundary node $2$ is closer
to the backbone node than boundary node $1$, $\alpha_2>\alpha_1$,
i.e., boundary node $2$ can help relay fewer packets for backbone
node $1$ before being rewarded. The two curves for $\alpha_1$ and
$\alpha_2$ for the same boundary node $1$ location cross at the
boundary node $1$ location. The figure shows that the average
fairness using the Shapley function gives greater rewards to the
boundary node whose channel is better and who can help the
backbone node more. When boundary node $2$ moves from $(20m,0m)$
to $(50m,0m)$, $\alpha_1$ becomes smaller, but $\alpha_2$ becomes
larger. This is because the backbone node must depend on boundary
node $2$ more for relaying. However, the backbone node will pay
less for the boundary nodes. Notice that $\alpha_i$ at the
crossover point is lower. This is because the overall power for
the backbone node is high when boundary node $2$ is far away, as
shown in Figure \ref{P0_R}.

\begin{figure}
    \centering
    \includegraphics[width=80truemm]{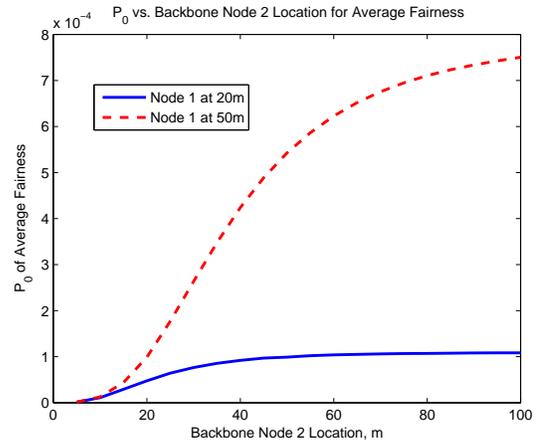}
    \caption{\footnotesize{$P_0$ of Average Fairness for Different Users' Locations}}
    \label{P0_shapley}
\end{figure}

\begin{figure}
    \centering
    \includegraphics[width=80truemm]{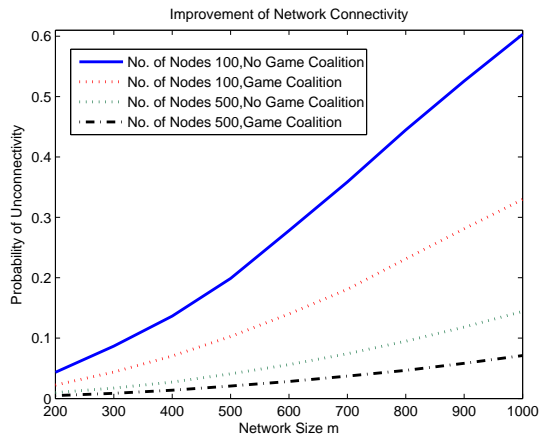}
    \caption{\footnotesize{Network Connectivity vs. Network Size}}
    \label{prob_size}
\end{figure}

Finally, we examine the degree to which the coalition game can
improve the network connectivity. Here we define the network
connectivity as the probability that a randomly located node can
connect to the other nodes. All nodes are randomly located within
a square of size $B\times B$. In Figure \ref{prob_size}, we show
the network un-connectivity as a function of $B$ for the numbers
of nodes equal to $100$ and $500$. With increasing network size,
the node density becomes lower, and more and more nodes are
located at the boundary and must depend on the others for
packet-forwarding. If no coalition game is formed, these boundary
nodes cannot transmit their packets due to the selfishness of the
other nodes. With the coalition game, the network connectivity can
be improved by about 50\%. The only chance that a node cannot
connect to the other nodes is when this node is located too far
away from any other node. We can see that the game coalition cures
the curse of the boundary nodes in wireless packet-forwarding
networks with selfish nodes.

% ------------------------------ SECTION ----------------------------------
\section{Conclusions}\label{sec:conclusion}

In this paper, we have proposed a coalition game approach to
provide benefits to selfish nodes in wireless packet-forwarding
networks using cooperative transmission, so that the boundary
nodes can transmit their packets effectively. We have used the
concepts of coalition games to maintain stable and fair game
coalitions. Specifically, we have studied two fairness concepts:
min-max fairness and average fairness. A protocol has been
constructed using repeated games and coalition games. From
simulation results, we have seen how boundary nodes and backbone
nodes form coalitions according to different fairness criteria. We
can also see that network connectivity can be improved by about
50\%, compared to the pure repeated game approach.

% ------------------------------ BIBLIOGRAPHY ----------------------------------
\bibliographystyle{IEEE}

\end{document}